\pdfoutput=1
\documentclass{aip-cp}

\usepackage[numbers]{natbib}
\usepackage{rotating}
\usepackage{graphicx}

\def\ket#1{|#1\rangle}

\newcommand{\ba}{\begin{eqnarray}}
\newcommand{\ea}{\end{eqnarray}}
\newcommand{\be}{\begin{equation}}
\newcommand{\ee}{\end{equation}}
\newcommand{\bmath}{\begin{mathletters}}
\newcommand{\emath}{\end{mathletters}}
\newcommand{\ban}{\begin{eqnarray*}}
\newcommand{\ean}{\end{eqnarray*}}

\newcommand{\bsub}{\begin{subequations}}
\newcommand{\esub}{\end{subequations}}

\begin{document}

\title{Excited-State Quantum Phase Transitions in Bosonic Lattice Systems}

\author{Michal Macek}

\affil{The Czech Academy of Sciences, Institute of Scientific Instruments, Brno, 612 64, Czech Republic}
\corresp{michal.macek@isibrno.cz}

\maketitle

\begin{abstract}
Concentrating on bosonic lattice systems, we ask whether and how Excited State Quantum Phase Transition (ESQPT) singularities occur in condensed matter systems with ground state QPTs. We study in particular the spectral singularities above the ground-state phase diagram of the boson Hubbard model. As a general prerequisite, we point out the analogy between ESQPTs and van Hove singularities (vHss). 
\end{abstract}

\bigskip

{\it INTRODUCTION: }  
Already in 1953, L. van Hove noticed singularities in the spectra of crystal vibrations related to the saddle points of the energy dispersion relations~\cite{ref:vanHove53}. Soon afterward, it was realized that similar effects occur in many different condensed-matter settings, prominently in electronic spectra, see~\cite{ref:Basani,ref:ElectronTopol} for reviews. The instability related to crossing of the van Hove singularity (vHs) with the Fermi energy may trigger phase changes in the systems, as described first by Lifshits~\cite{ref:Lifshits}. 
Recently, similar level-density singularities have been observed in ``zero-dimensional'' interacting many-body systems, like atomic nuclei and molecules, and were described in terms of excited state quantum phase transitions (ESQPTs)~\cite{ref:Cejn06,ref:Capr07}. These generalize the concept of quantum phase transitions (QPTs)~\cite{ref:Carr} to excited spectra, and have currently attracted considerable attention in studies of atomic nuclei, molecules, coupled atom-field quantum optics systems, driven quantum oscillators, and two dimensional lattices. The interest in ESQPTs stems on one hand from the marked structural changes occurring in the individual systems at critical excitation energies, on the other hand from possible profound general implications for non-equilibrium thermodynamics, quantum information processing, and transport. As in standard QPTs, any non-thermal quantities that can be varied in the system's Hamiltonian can serve as the control parameters $\lambda$. Examples are the number of nucleons in atomic nuclei, the pump power or the pump-cavity detuning in atom field systems, or the strength of the driving force in driven systems. For overview of literature on various aspects of ESQPTs, see~\cite{ref:CejnHERE}. 

In this contribution, we address the question whether and how ESQPT singularities occur in condensed matter systems with ground state QPTs. As a prerequisite, we point out the analogy between ESQPTs and vHss: ESQPTs occur as non-analytic points of the semi-classical energy level density $\overline{\rho}_\lambda(E)$ at critical values of excitation energy $E = E_\mathrm{ESQPT} > E_\mathrm{gs}$, which in general evolve with the control parameters $\lambda$ of the system, i.e. $E_\mathrm{ESQPT} = E_\mathrm{ESQPT}(\lambda)$,~\cite{ref:Capr07,ref:esqptI,ref:esqptII,ref:esqptPLA}. Considering the definition of the energy level density [usually termed ``density of states'' (DoS) in condensed matter context]:
\ba\label{eq:DOS}
\overline{\rho}^\lambda(E) \propto \int\dots\int d^f\vec{p}\,d^f\vec{x}\, \delta[E-H^\lambda_{\mathrm{cl}}(\vec{p},\vec{x})]\,\quad\leftrightarrow\quad \overline{\mathrm{DoS}}^\lambda(E) &\propto& \int\dots\int d^{\tilde{f}}\vec{k}\, \delta[E-E^\lambda(\vec{k})]\,,
\ea
where one can immediately notice that the classical Hamiltonian, $H^\lambda_{\mathrm{cl}}(\vec{p},\vec{x})$, and the energy dispersion relation (EDR), $E^\lambda(\vec{k})$, play analogous roles. The only difference lies in the dimensionalities $f$ and $\tilde{f}$.
In spatially extended systems, the DoS is expressed by the integral over the $\tilde{f}$-component quasimomenta $\vec{k}$ in $E^\lambda(\vec{k})$. Here, the number of freedom degrees $\tilde{f} = \nu\cdot D$ is determined by the number of particles $\nu$ and the dimension of space $D$.
In spatially zero-dimensional systems with $f$ internal degrees of freedom, the level density integral is taken over the two $f$-component vectors of the canonical coordinates $\vec{x}$ and the conjugate momenta $\vec{p}$ available at energy $E$; the dimension is even. Thus equivalent types of singularities may appear if $\tilde{f}=2f$ and if $H^\lambda_{\mathrm{cl}}(\vec{p},\vec{x})$ and $E^\lambda(\vec{k})$ possess equivalent types of stationary points (saddle, local maximum, ...), see~\cite{ref:CejnHERE,ref:esqptI,ref:esqptPLA}.
Considering the above analogy, we explore here the occurrence of the ESQPT spectral singularities of the boson Hubbard (BH) model on a 1D lattice, ``above'' the phase diagram of the model~\cite{ref:Monien}. Let us note that the BH model is equivalent to an $N\rightarrow\infty$ contraction of the coupled U(2) Vibron model~\cite{ref:IachLie}, and the study presented here may naturally be extended within the rich family of Vibron models~\cite{ref:IachLev}.  
\begin{figure}[t]\label{fig:1}
  \centerline{\includegraphics[width=200pt]{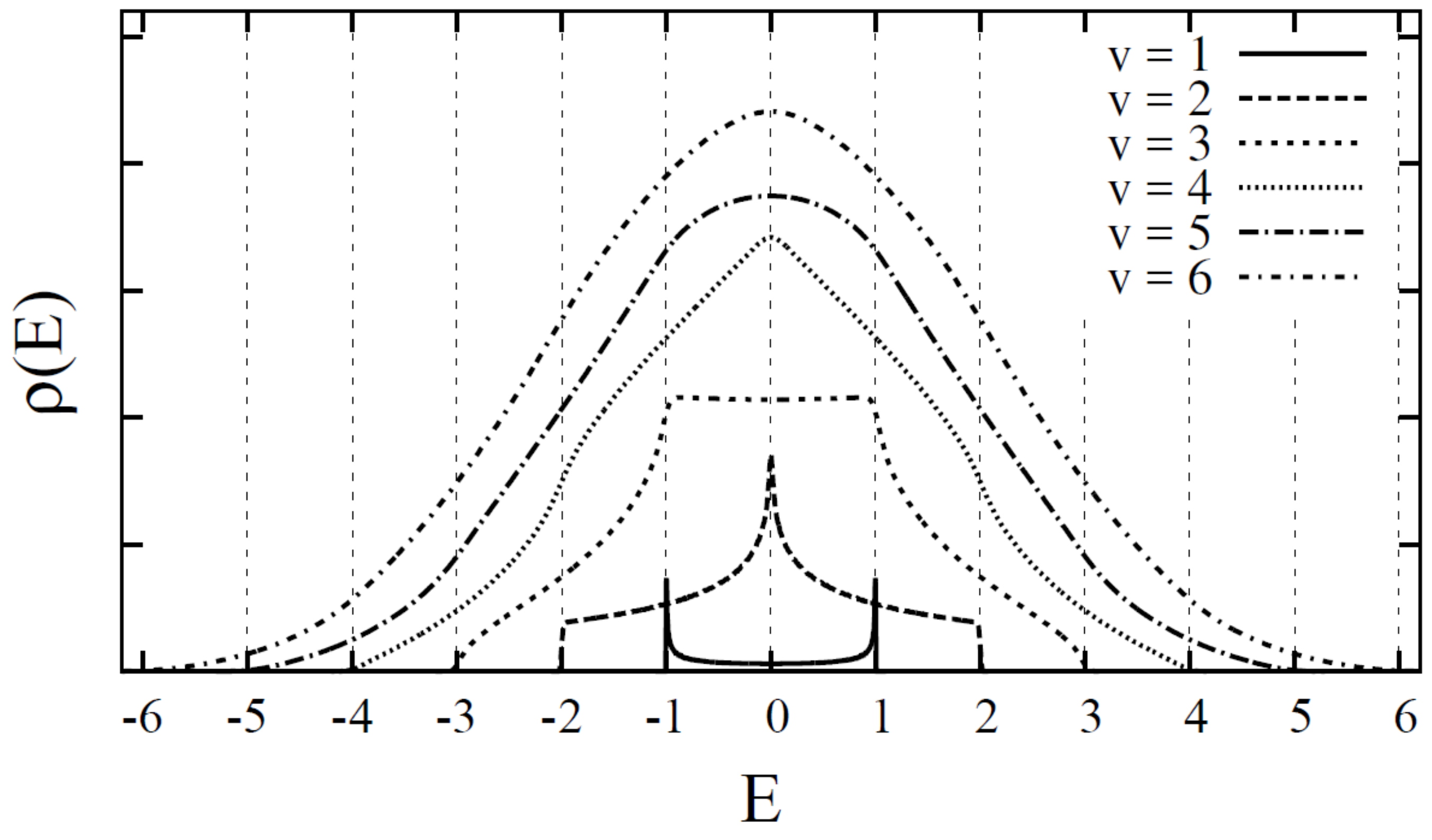}}
  \caption{Level densities of the hopping limit ($U=V=0$) of the Bose-Hubbard Hamiltonian for $\nu = 1,2,...,6$ particles. Inverse square-root singularities are seen for $\nu=1$ at band-edges at $E = \pm 1$ and a logarithmic singularity is seen for $\nu = 2$ at mid-band $E = 0$. For $\nu > 2$, the singularities affect higher $E$-derivatives of $\rho(E)$.}
\end{figure}



{\it 1D BOSON HUBBARD MODEL:}
The Hamiltonian of the boson Hubbard model is:
\ba\label{eq:HBH}
H = -t \sum_{\langle i,j\rangle}^{\nu} \hat{M}_{ij} -\mu\sum_{i=1}^\nu \hat{n}_i + U/2\sum_{i=1}^\nu\hat{n}_i(\hat{n}_i - 1) + V \sum_{\langle i,j\rangle}^{\nu}\hat{n}_i\hat{n}_j\,,
\ea
where $\hat{n}_i = b^\dag_i b_i$ is the boson number operator at site $i$, and $M_{ij} = (b^\dag_i b_j + b^\dag_j b_i)$ is the inter-site hopping operator (equivalent to the harmonic limit of the Majorana operator in the coupled U(2) Vibron model), considered here to act between nearest neighbor (NN) sites $\langle i,j\rangle$ on a linear lattice of length $L$. The coefficients $t,\mu,U,V$ serve as control parameters for the (ES)QPTs. 

Eigenstates corresponding to total number of $\nu$ bosons take in the local limit $t=0$ the form $\ket{\underbrace{1...1}_\nu\underbrace{0...0}_{L-\nu}}$ (so-called ``combination modes''), $\ket{2\underbrace{1...1}_{\nu-1}\underbrace{0...0}_{L-\nu}}$ (first overtones), etc... For $t\neq 0$ and $U=V=0$, the particles delocalize and the $\nu$-particle dispersion relation is $E^t_\nu = -t\sum_{\alpha=1}^\nu \cos[\pi k_\alpha/(L+1)]$, where $k_\alpha$ represent the quasimomenta of individual particles. The spectra for $\nu = 1, 2, ..., 6$-particle bands of the ``free-hopping'' limit ($U = V = 0$) are shown for $t=1$ and $\mu = 0$ in Fig. 1. Notice the singularities, apparent at energies $E_s$, related to the stationary points of the dispersion relation $E^t_\nu$: In case of $\nu = 1$, there are $\overline\mathrm{DoS}(E)\approx 1/\sqrt{|E - E_s|}$-singularities at both band-edges $E_s = \pm 1$ and for $\nu = 2$ there is a $\overline\mathrm{DoS}\approx \log{|E-E_s|}$ at mid-band $E_s=0$. Bands with $\nu = 3$ and $4$ contain ``inverse-square-root'' and logarithmic singularities in the derivative $\partial \overline\mathrm{DoS}/\partial E$, respectively. In general, the singularities occur in derivatives of order $\lfloor(\nu-1)/2\rfloor$, and are of inverse-square-root (logarithmic) type for odd (even) $\nu$, cf.~\cite{ref:esqptPLA}. 

\begin{figure}[h]\label{fig:2}
  \centerline{\includegraphics[width=\linewidth]{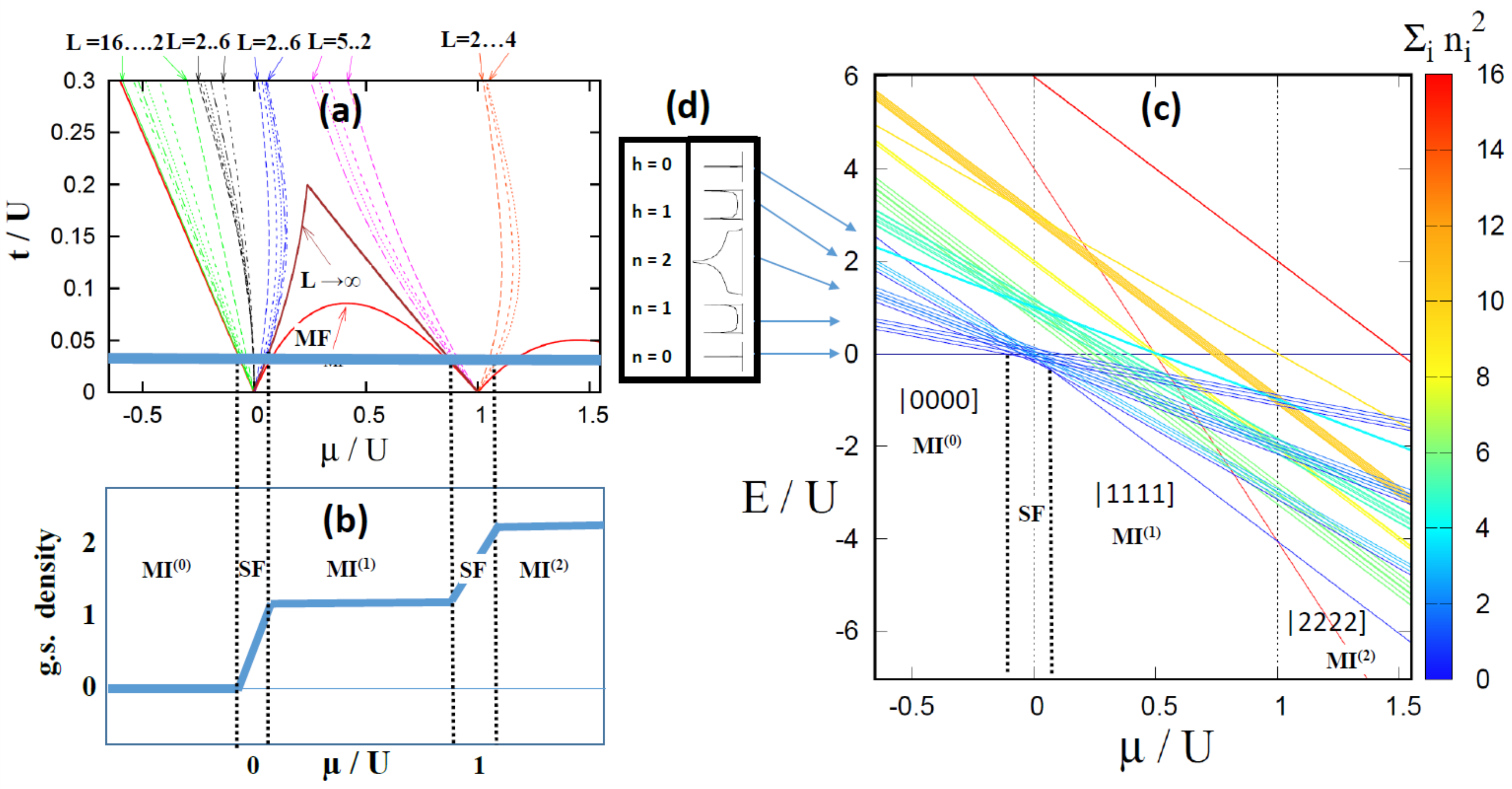}}
  \caption{Panel (a): Phase diagram of the BH Hamiltonian, Eq. (2), containing the SF phase and the insulator phases MI$^{(0)}$, MI$^{(1)}$, MI$^{(2)}$. Panel (b): Evolution of the order parameter---the g.s. density for $t/U = 0.035$. Panel (c): BH spectrum for $L = 4$ sites and $n = 0,1,2,3,4$ particles (additional $n=8$, $|2222\rangle$ level is shown), the color-code expresses the expectation value $\langle \sum_i n_i^2\rangle$ corresponding to individual levels. Panel (d): Schematic level density for the $n=0,1,2$-particle and $h=0,1$-hole bands in the limit $L=\rightarrow\infty$. Arrows indicate the relation to the bands shown in panel (c).}
\end{figure}

Figure 2~(a) represents the phase diagram of the BH model, showing the transition lines between the superfluid (SF) phase and the Mott insulator phases MI$^{(0)}$, MI$^{(1)}$, MI$^{(2)}$,... (with ground state densities $\nu_\mathrm{g.s.}/L = 0, 1,2,...$). The phase boundaries are shown here for (i) mean field (MF), cf.~\cite{ref:Sach}, (ii) NN interactions with $L\rightarrow \infty$, cf.~\cite{ref:Monien}, and (iii) NN interactions for several finite lattice lengths $L$. 
Panel (b) shows the order parameter---the g.s. density---in the $L\rightarrow\infty$ limit corresponding to $t/U = 0.035$ as a function of $\mu/U$, reflecting a phase transition that is continuous for $t > 0$. 
Panel (c) shows the spectrum of $\nu = 0,1,2,3,4$ levels as a function of $\mu/U$ for $t/U = 0.035$ fixed. For $\mu < -t$, the ground state is the $\nu = 0$ particle vacuum $\ket{0000}$, as appropriate for the MI$^{(0)}$ Mott insulator phase, while for $\mu > t$ it is the $\nu = 4$ ``hole vacuum'' $\ket{1111}$ corresponding to MI$^{(1)}$ Mott insulator with one particle per site density. 
In the superfluid (SF) phase, between $-t < \mu < t$, the nature of the g.s. changes due to crossings of the bands with $\nu=1,2,3$, which contain $L, L(L-1)/2$ and $L$ states, respectively. Notice that for $L = 4$, $\nu=2$ corresponds to half-filling situation, while $\nu=3$ is a one-hole band [denoted alternatively ``$h=1$'' in Fig. 2~(c)]. To first order in $t/U$, these bands are composed of the combination modes only. 
ESQPT singularities  of the types shown in Fig. 1 appear in individual bands in the $L\rightarrow\infty$ limit. The singularities are sharp in bands with low numbers of particles, or holes, as indicated in Fig. 2~(d), while the density of the half-filling band approaches a Gaussian. 
As in the case of topological phase transitions, ESQPTs can not be simply related to phases via order parameters in the sense of Landau. Possible approaches include considering energy derivatives of average values of suitable operators at given energy, which distinguish different response of the system to perturbations in the different phases, as done in Ref.~\cite{ref:Cejn16}, or defining order parameters via quasi dynamical symmetries~\cite{ref:Mace15}. ESQPT phases, nature of singularities in higher lying bands with contribution of overtones, and more complex types of lattices~\cite{ref:Crystals} will be subject of further study.

I am deeply thankful to Franco Iachello for broad inspiration and chances to learn in many areas of not only physics, in particular for initiating this project and patient guidance during my postdoctoral stay with him at Yale. 
The work is supported by MEYS CR project CZ.02.2.69/0.0/0.0/18 070/0009944 (qCULTURA).



\nocite{*}
\bibliographystyle{aipnum-cp}%

\end{document}